\documentclass[]{spie}  

\usepackage[utf8]{inputenc}
\usepackage{amsmath,amsfonts,amssymb}
\usepackage{graphicx}
\usepackage[colorlinks=true, allcolors=blue]{hyperref}
\usepackage{upgreek}
\usepackage{epstopdf}
\usepackage{epsfig}
\usepackage{float}

\usepackage{siunitx}

\usepackage{acronym}
\usepackage[caption=false]{subfig}

\acrodef{AO}[AO]{adaptive optics}
\acrodef{CCD}{charged-coupled device}
\acrodef{DM}[DM]{deformable mirror}
\acrodef{FLAO}{First Light AO}
\acrodef{ISYS}[ISYS]{Institute for System Dynamics Stuttgart}
\acrodef{KIT}[KIT]{Karlsruhe Institute for Technology}
\acrodef{KOOL}[KOOL]{Koenigstuhl Observatory Opto-Mechatronics Laboratory}
\acrodef{LBT}[LBT]{Large Binocular Telescope}
\acrodef{LSW}[LSW]{Landessternwarte}
\acrodef{MCF}[MCF]{multi-core fiber}
\acrodef{MFD}[MFD]{mode-field diameter}
\acrodef{MLA}[MLA]{micro-lens array}
\acrodef{MLR-TT}[MLR-TT]{microlens ring tip-tilt sensor}
\acrodef{MM}[MM]{multi-mode}
\acrodef{MMF}[MMF]{multi-mode fiber}
\acrodef{MPIA}[MPIA]{Max Plank Institute for Astronomy}
\acrodef{NA}[NA]{numerical aperture}
\acrodef{NCP}[NCP]{non-common path}
\acrodef{NAIR}[NAIR]{Novel Astronomical Instrumentation based on photonic light Reformating}
\acrodef{NIR}[NIR]{near-infrared}
\acrodef{PSF}[PSF]{point-spread function}
\acrodef{PSD}[PSD]{power spectral density}
\acrodef{SM}[SM]{single-mode}
\acrodef{SMF}[SMF]{single-mode fiber}
\acrodef{TPP}[TPP]{two-photon polmerization}
\acrodef{WFS}[WFS]{wavefront sensor}
\acrodef{ExAO}[ExAO]{extreme adaptive optics}

\title{Long-term performance of the MLR-TT sensor: two-photon polymerization validation in the telescope environment}

\author[a]{Robert J. Harris}
\author[]{Philipp Hottinger}
\author[b]{Jonathan Crass}
\author[c]{Philipp-Immanuel Dietrich}
\author[d,e]{Christian Koos}

\affil[a]{Centre for Advanced Instrumentation, Department of Physics, Durham University, South Road, Durham, DH1 3LE, UK}
\affil[b]{Department of Astronomy, The Ohio State University, 4055 McPherson Laboratory, 140 West 18$^{\mathrm{th}}$ Ave., Columbus, OH 43210, USA}
\affil[c]{Vanguard automation GmbH, Hermann-von-Helmholtz-Platz 1, 76344~Eggenstein-Leopoldshafen, Germany}
\affil[d]{Institute of Microstructure Technology~(IMT), Karlsruhe Institute of Technology~(KIT), Hermann-von-Helmholtz-Platz~1, 76344~Eggenstein-Leopoldshafen, Germany}
\affil[e]{Institute of Photonics and Quantum Electronics~(IPQ), Karlsruhe Institute of Technology~(KIT), Engesserstr.~5, 76131~Karlsruhe, Germany}

\authorinfo{Further author information: (Send correspondence to R.J.H)\\P.H.: E-mail: robert.j.harris@durham.ac.uk}

\begin{document}
\maketitle

\begin{abstract}

We report on the long-term optical performance of a two-photon polmerized \acl{MLR-TT}. The study is backed by repeated measurements over the last six years, since the sensor was first tested in the sky at the Large Binocular Telescope. The goal of the study is to assess the feasibility of the underlying technology of two-photon polymerization for future instrumentation in the realistic environments experienced at astronomical telescopes.

\end{abstract}

\keywords{single-mode fiber, fiber coupling, tip-tilt sensor, two-photon polymerization, spectroscopy, micro-lens array, }


\section{Introduction}

\Ac{TPP} is an additive manufacturing technique that enables fabrication on nano- and micro-scales \cite{Dietrich2018}. It has been used to produce structures for a wide range of applications, spanning fields from medicine to engineering \cite{OHalloran2023}. Of particular interest to astronomy is its ability to create custom, lightweight, and highly precise optical components that are difficult, or in some cases impossible, to manufacture using conventional techniques \cite{Dietrich2018}. 

For \ac{TPP} structures to be adopted in astronomical instrumentation, however, their performance and reliability must be demonstrated over timescales comparable to the operational lifetime of an instrument. Several studies have investigated their behaviour under laboratory conditions that emulate harsh observational environments, including cryogenic temperatures and vacuum operation \cite{Peek2022,Ruchka_2022}. While these tests are encouraging, practical deployment at observatories requires evidence that such components can maintain their performance during long-term operation at a telescope site. 

\begin{figure}
    \centering
    \includegraphics[width=0.5\linewidth]{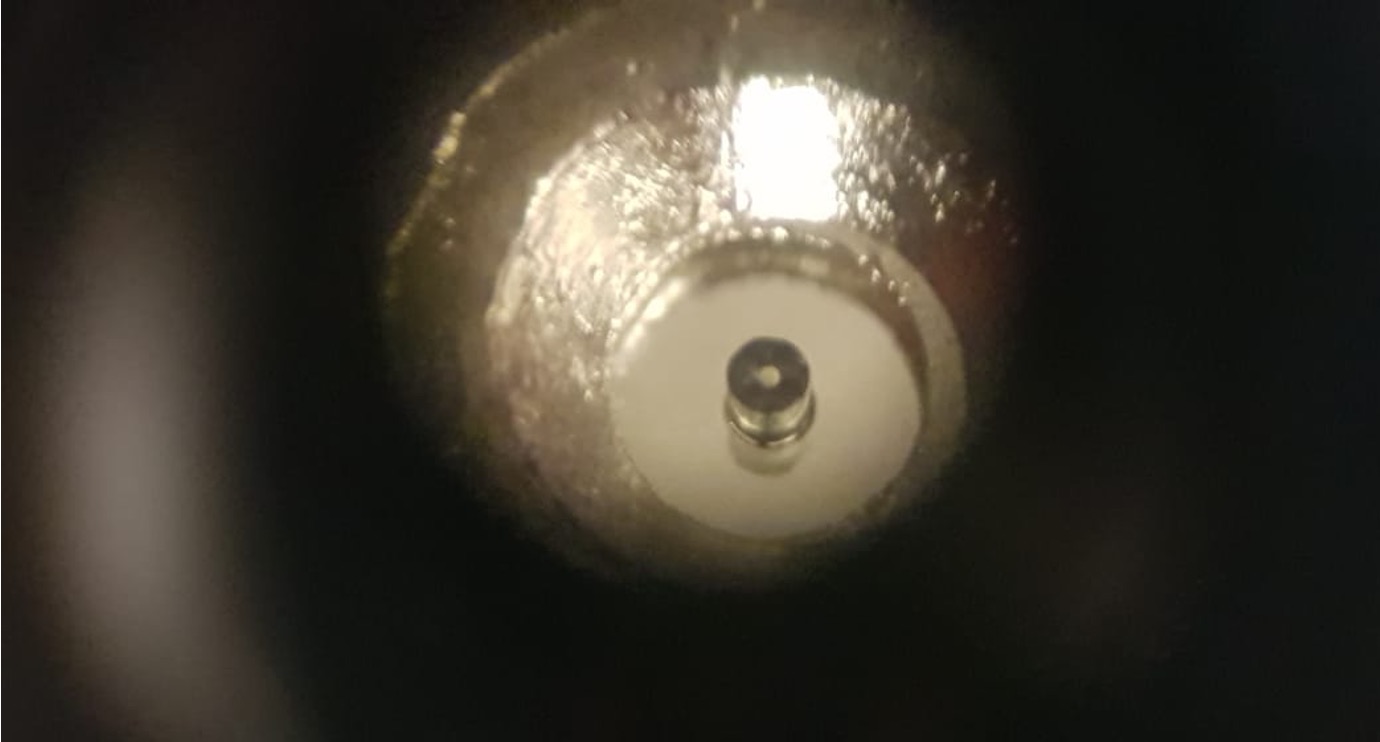}
    \caption{A camera image of the MLR-TT as installed in the iLocater acquisition camera. The MLR-TT was created using two-photon polymerization. The ring contains six microlenses and has a diameter of 355\si{\micro\meter} and stands 400\si{\micro\meter} high.}
    \label{fig:MLT_TT_in_situ}
\end{figure}

In this work, we present the long-term performance of the \ac{MLR-TT}, a \ac{TPP}-fabricated sensor (Figure~\ref{fig:MLT_TT_in_situ}) that has been installed at the \ac{LBT} since November 2019 \cite{Hottinger21}. We report the results of a series of optical tests conducted over six years following installation. To our knowledge, this represents the longest continuous in-situ telescope deployment and performance assessment of a two-photon-polymerized optical component.

In section \ref{sec:methods} we describe the MLR-TT sensor and our experiment. In section \ref{sec:results} we show the results of our tests. In section \ref{sec:discussion} we discuss what this means and future pathways and we conclude in section \ref{sec:conclusions}.

\section{Methods}
\label{sec:methods}

\subsection{MLR-TTS}

\begin{figure}
    \centering
    \begin{tabular}{cccc}
     \includegraphics[width=0.22\linewidth]{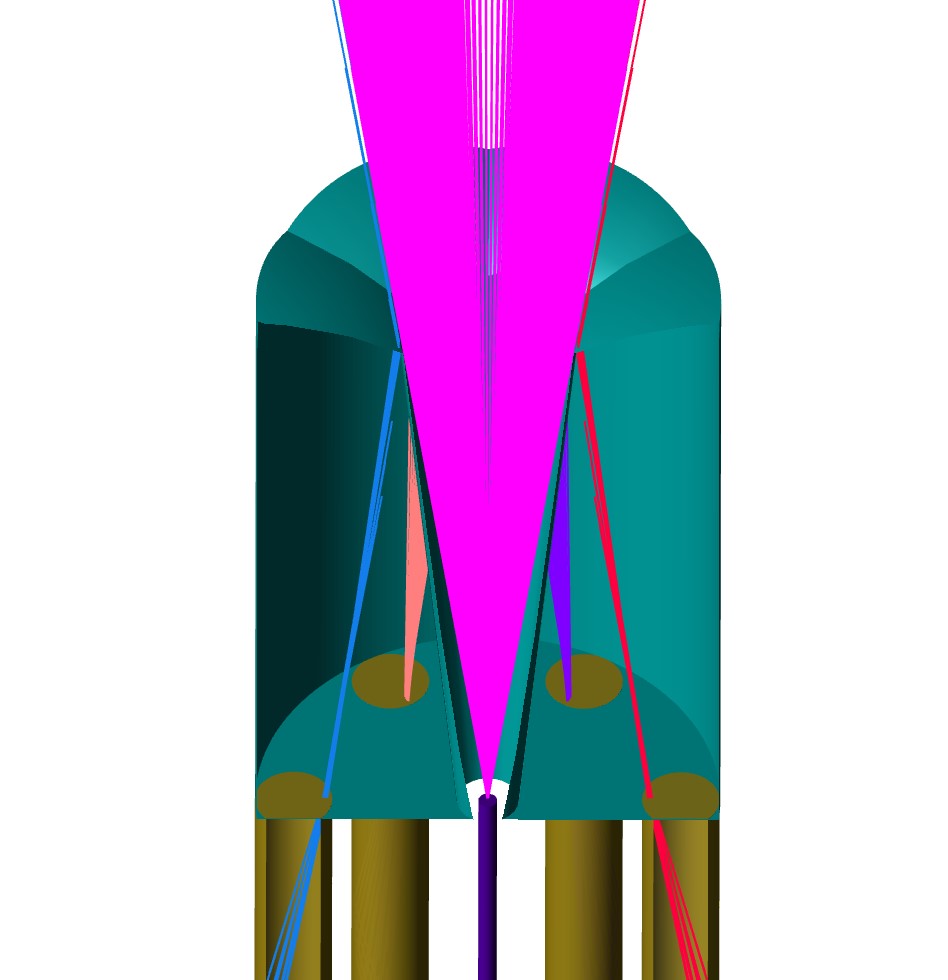}    & \includegraphics[width=0.22\linewidth]{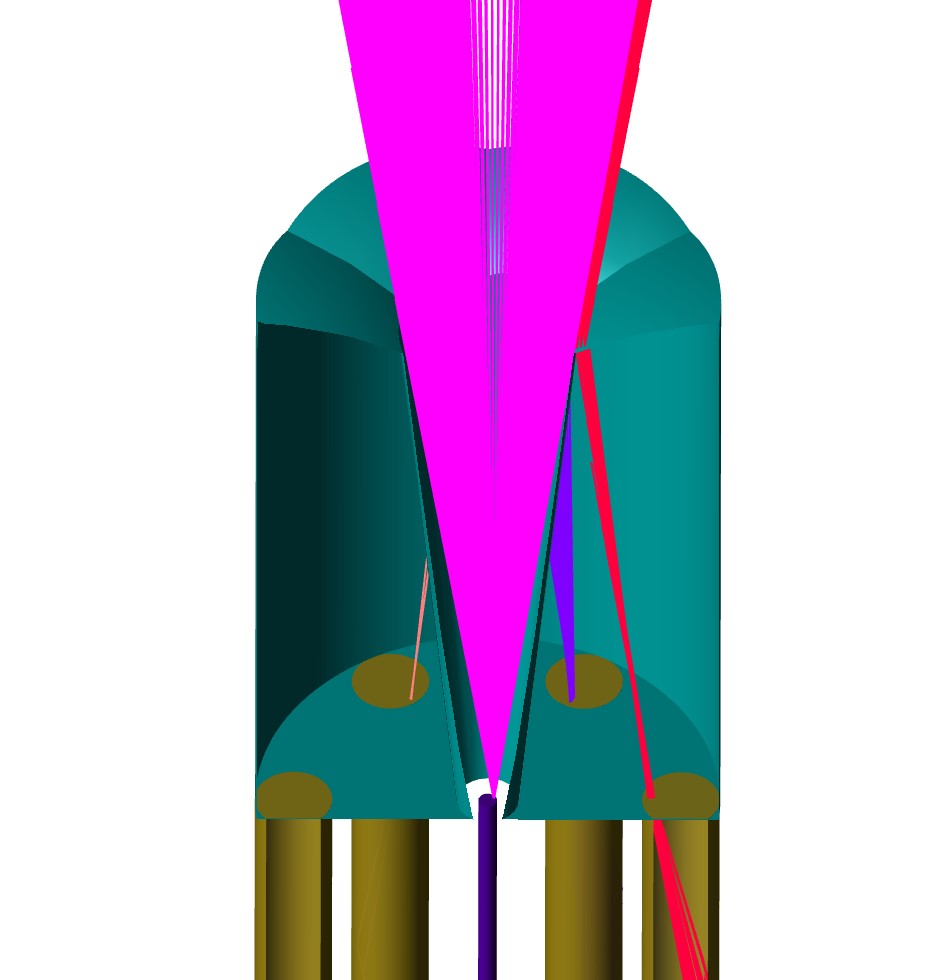} & \includegraphics[width=0.22\linewidth]{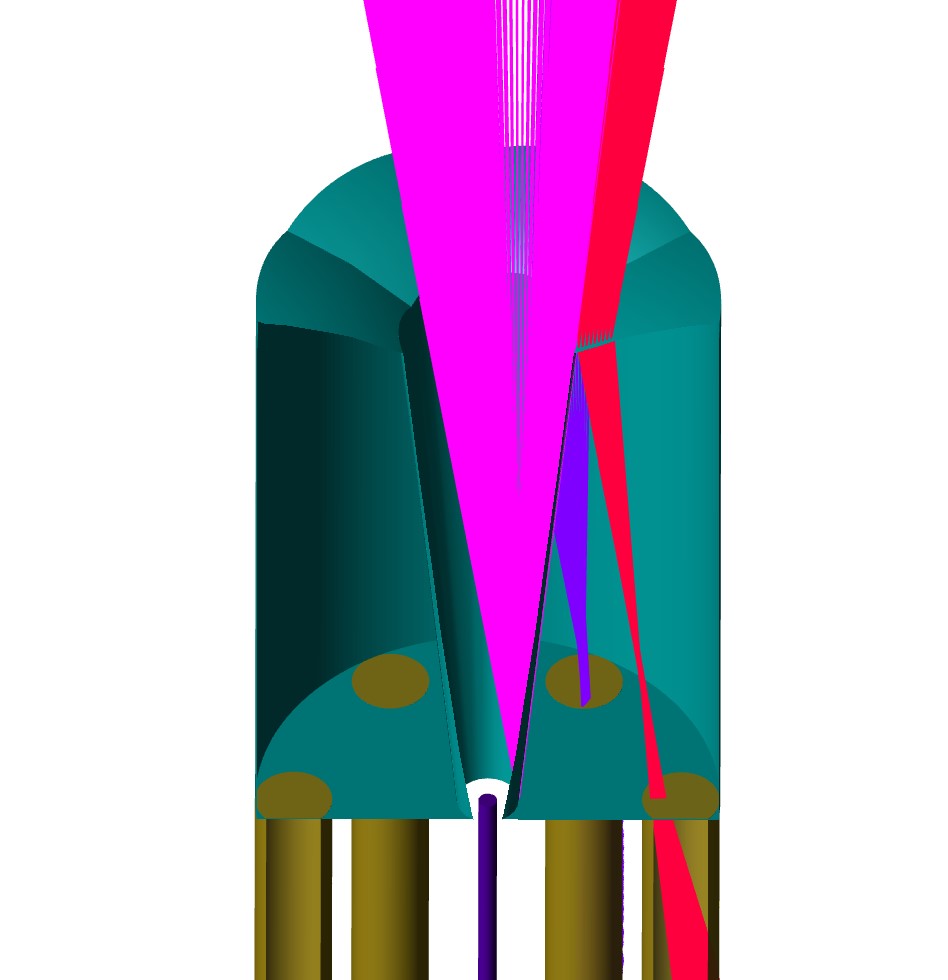} &
     \includegraphics[width=0.22\linewidth]{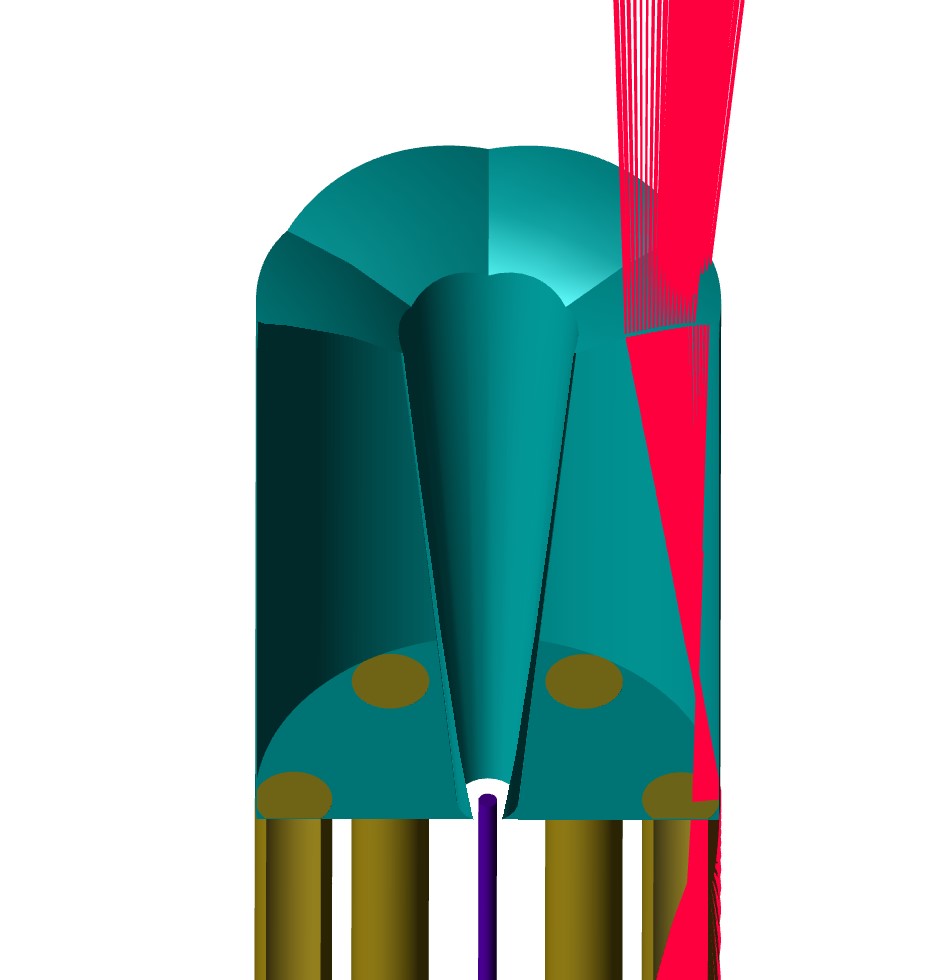}\\     
    \end{tabular}
    \caption{Schematic of the MLR-TT sensor. On the left we show the aligned sensor, with the majority of the light coupling into the central single mode science fiber. In the left of middle the beam is misaligned by $\sim$2\,\si{\micro\meter}, increasing coupling to the multimode fibers. On the right of middle the beam is misaligned by $\sim$10\,\si{\micro\meter}, with no light coupling to the single mode fiber. On the right hand side the beam is misaligned by $\sim$50\,\si{\micro\meter} and is coupled to one multimode fiber. Reproduced from Hottinger+2018\cite{hottinger2018}.}
    \label{fig:MLR_tts_response}
\end{figure}

The \ac{MLR-TT} is designed to correct residual tip-tilt errors that are not sensed by the \ac{LBT} \ac{AO} system, while enabling simultaneous injection of science light into a fiber and wavefront sensing at the same focal plane. The sensor consists of a central single-mode fiber that carries the science light, surrounded by six multimode fibers used for tip-tilt sensing (Figure~\ref{fig:MLR_tts_response}). A \ac{TPP}-fabricated microlens ring is mounted above the six multimode fibers to enhance the wavefront-sensing sensitivity. The microlens ring is designed to direct approximately 13\% of the incident on-axis light into the multimode fibers, where it is used for guiding. As the beam undergoes tip or tilt displacements, the flux distribution among the multimode fibers changes, providing a measurement of the pointing error. This redistribution of light comes at the expense of science throughput, reducing the theoretical coupling efficiency of the central fiber from approximately 80\% to 65\%. Further details of the \ac{MLR-TT} and original experiment can be found in Hottinger et al. 2018\cite{hottinger2018} and Hottinger et al. 2021 \cite{Hottinger21}.

\subsection{Experimental setup}

The MLR-TT has been left in the iLocater acquisition camera since November 2019 \cite{crass2019}. The front end is a closed box, but is exposed to the telescope environment, with fluctuations in humidity, temperature along with the vibrations and movements of the telescope. This provides a realistic environment for our tests.
Over the last six years, the iLocater acquisition camera and associated optics have undergone several upgrades. This means the light sources and alignment, along with the \ac{PSF} have changed on several occasions making absolute comparisons between measurements impossible, thus we only test the relative response of the MLR-TT.
We use the instrument internal mechanisms and calibration sources, including a five-axis stage for fiber coupling to maximize the light entering the science single-mode fiber. We use a controllable tip/tilt mirror to perform a circular scan of the calibration source beam across the micro-lens focal plane to examine the response of the MLR-TT in each of the multimode fibers.

\begin{figure}
    \centering
    \includegraphics[trim={0 3cm 0 .5cm},clip,width=0.5\linewidth]{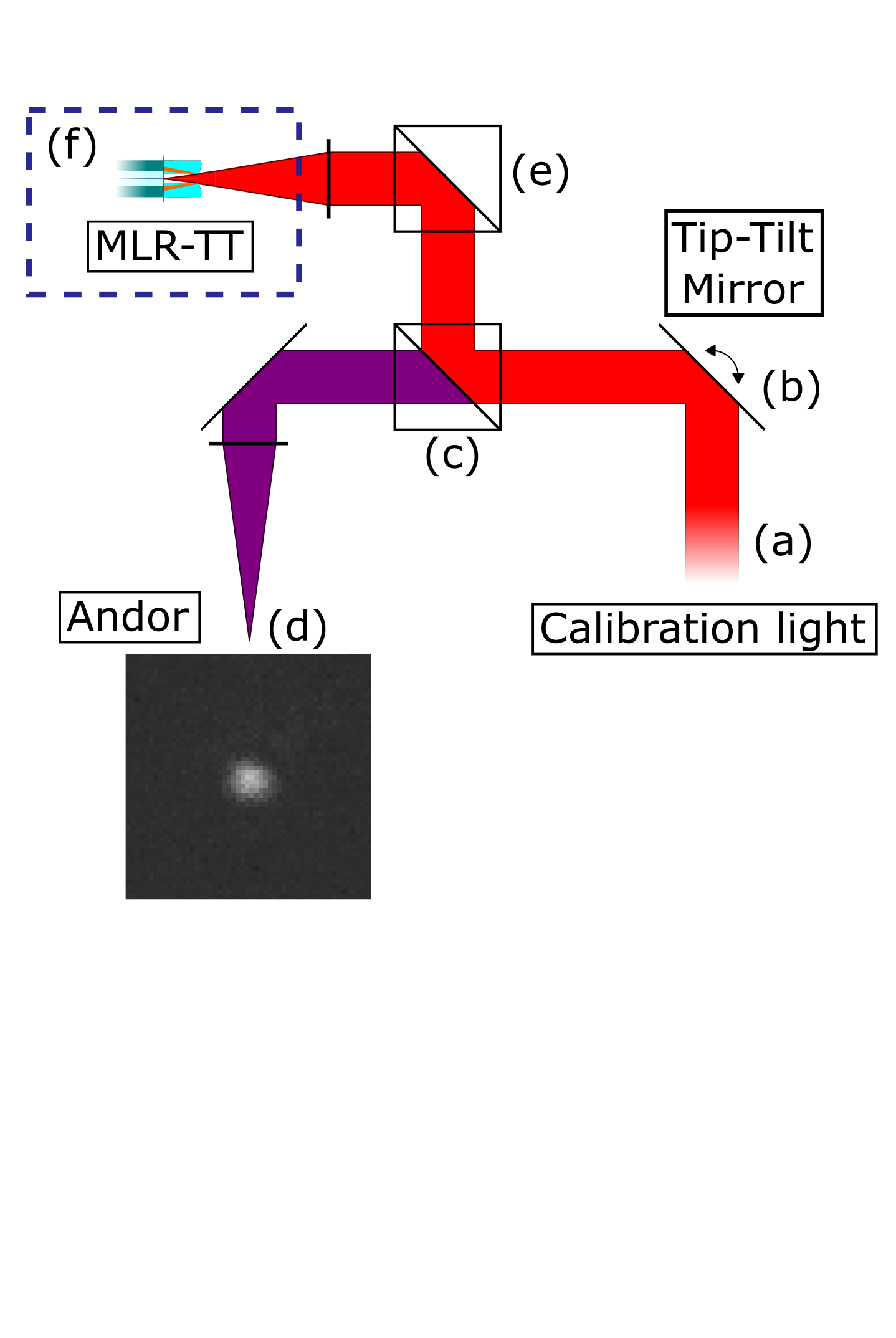}
    \caption{Schematic of the experimental setup. Here collimated calibration light (a) is supplied into the system. The tip-tilt mirror (b) allows beam steering. The dichrioic (c) picks off short wavelength light for imaging on the Andor camera whilst the longer wavelength light is sent through the second dichroic (e) to the MLR-TT (f).}
    \label{fig:placeholder}
\end{figure}

\section{Results}
\label{sec:results}

\begin{figure}
    \centering
    \includegraphics[width=1\linewidth]{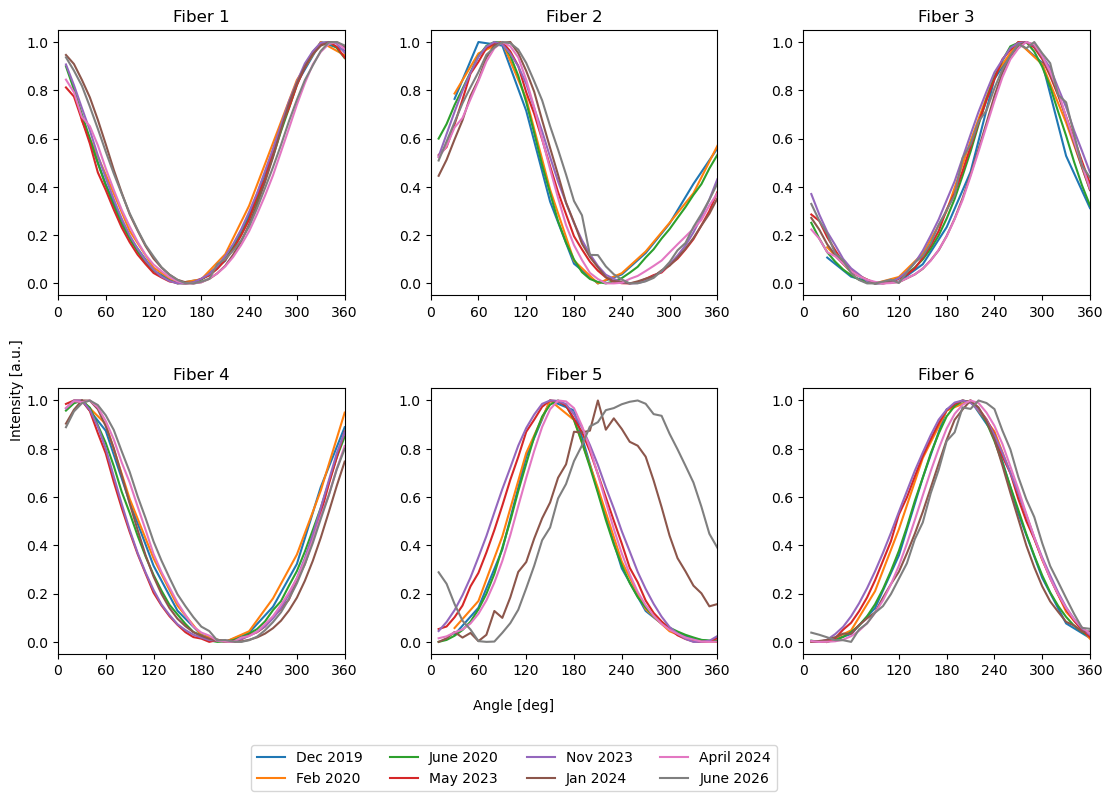}
    \caption{The measured output intensity at each multimode fiber during circular scans of the incident beam. The dates of the scans can be found in the legend, at the bottom. Note that the shape of the curves is very similar, but the tests in Jan 2024 and June 2026 show a phase shift for fibers 2 and 5.}
    \label{fig:MLT_TT_response}
\end{figure}

We plot the results from the circular scans taken over several years in Figure \ref{fig:MLT_TT_response}. All results are normalized to account for differing flux from the differing light sources used to perform the scan. Importantly, they show similar response between all fibers and tests. Notably the phase for May 2023 and June 2026 in fibers 2 and 5 is shifted compared to the rest of the tests. We ascribe this to a misalignment in one axis with respect the iLocater acquisition camera.

\section{Discussion}
\label{sec:discussion}

The results presented indicate that the MLR-TT sensor is still attached to its fiber and functioning as in 2019. Although changes to the acquisition camera prevent absolute throughput comparisons between epochs, the relative response functions remain  consistent. We note the variations in the phase response for two tests, likely due to misalignment with respect to the iLocater acquisition camera.

Overall, the results indicate  that possible degredation mechanisms at the \ac{LBT}: humidity, thermal variations, contamination and vibrations; do not appear to have affected the sensor performance. A full validation of the system will be conducted once the MLT-TT sensor has been removed from the telescope, which is planned for later in 2026.

\section{Conclusions}
\label{sec:conclusions}

The \acl{MLR-TT} sensor, which has been left at the \acl{LBT} for the last six years shows consistent performance across this time period and is  a validation of the long-term viability of \acl{TPP}. These test measurements represent, to our knowledge, the longest telescope deployment of a two-photon-polymerized optical component. The stability observed after six years demonstrates that acl{TPP}-fabricated micro-optics can survive and operate reliably in observatory environments, supporting their use in future astronomical instrumentation.

We intend to continue testing the \acl{MLR-TT} during the iLocater commissioning, before shipping it to Durham for in-depth laboratory inspection and testing.

\acknowledgments

Robert J. Harris received support as part of the “NAIR-APREXIS” grant, funded by the Deutsche Forschungsgemeinschaft (grant no. 506421303).

\bibliography{phd}

@article{crass2019,
    author = {Crass, J and Bechter, A and Sands, B and King, D and Ketterer, R and Engstrom, M and Hamper, R and Kopon, D and Smous, J and Crepp, J R and Montoya, M and Durney, O and Cavalieri, D and Reynolds, R and Vansickle, M and Onuma, E and Thomes, J and Mullin, S and Shelton, C and Wallace, K and Bechter, E and Vaz, A and Power, J and Rahmer, G and Ertel, S},
    title = {Final design and on-sky testing of the iLocater SX acquisition camera: broad-band single-mode fibre coupling},
    journal = {Monthly Notices of the Royal Astronomical Society},
    volume = {501},
    number = {2},
    pages = {2250-2267},
    year = {2021},
    month = {02},
    issn = {0035-8711},
    doi = {10.1093/mnras/staa3355},
    url = {https://doi.org/10.1093/mnras/staa3355},
    eprint = {https://academic.oup.com/mnras/article-pdf/501/2/2250/35393727/staa3355.pdf},
}

@article{Dietrich2018,
archivePrefix = {arXiv},
arxivId = {1802.00051},
author = {Dietrich, P.-I. and Blaicher, M. and Reuter, I. and Billah, M. and Hoose, T. and Hofmann, A. and Caer, C. and Dangel, R. and Offrein, B. and Troppenz, U. and Moehrle, M. and Freude, W. and Koos, C.},
doi = {10.1038/s41566-018-0133-4},
eprint = {1802.00051},
isbn = {4156601801334},
issn = {1749-4885},
journal = {Nature Photonics},
month = {apr},
number = {4},
pages = {241--247},
publisher = {Springer US},
title = {{In situ 3D nanoprinting of free-form coupling elements for hybrid photonic integration}},
url = {http://dx.doi.org/10.1038/s41566-018-0133-4 http://www.nature.com/articles/s41566-018-0133-4},
volume = {12},
year = {2018}
}

@inproceedings{hottinger2018,
author = {Philipp Hottinger and Robert J. Harris and Philipp-Immanuel Dietrich and Matthias Blaicher and Martin Gl{\"u}ck and Andrew Bechter and Jonathan Crass and J{\"o}rg-Uwe Pott and Christian Koos and Oliver Sawodny and Andreas Quirrenbach},
title = {{Micro-lens arrays as tip-tilt sensor for single mode fiber coupling}},
volume = {10706},
booktitle = {Advances in Optical and Mechanical Technologies for Telescopes and Instrumentation III},
editor = {Ram{\'o}n Navarro and Roland Geyl},
organization = {International Society for Optics and Photonics},
publisher = {SPIE},
pages = {1070629},
year = {2018},
doi = {10.1117/12.2312015},
URL = {https://doi.org/10.1117/12.2312015}
}

@article{Hottinger21,
author = {Philipp Hottinger and Robert J. Harris and Jonathan Crass and Philipp-Immanuel Dietrich and Matthias Blaicher and Andrew Bechter and Brian Sands and Timothy Morris and Alastair G. Basden and Nazim Ali Bharmal and Jochen Heidt and Theodoros Anagnos and Philip L. Neureuther and Martin Gl\"{u}ck and Jennifer Power and J\"{o}rg-Uwe Pott and Christian Koos and Oliver Sawodny and Andreas Quirrenbach},
journal = {J. Opt. Soc. Am. B},
number = {9},
pages = {2517--2527},
publisher = {Optica Publishing Group},
title = {On-sky results for the integrated microlens ring tip-tilt sensor},
volume = {38},
month = {Sep},
year = {2021},
url = {https://opg.optica.org/josab/abstract.cfm?URI=josab-38-9-2517},
doi = {10.1364/JOSAB.421459},
}

@article{OHalloran2023,
author = {O'Halloran, Seán and Pandit, Abhay and Heise, Andreas and Kellett, Andrew},
title = {Two-Photon Polymerization: Fundamentals, Materials, and Chemical Modification Strategies},
journal = {Advanced Science},
volume = {10},
number = {7},
pages = {2204072},
doi = {https://doi.org/10.1002/advs.202204072},
url = {https://advanced.onlinelibrary.wiley.com/doi/abs/10.1002/advs.202204072},
eprint = {https://advanced.onlinelibrary.wiley.com/doi/pdf/10.1002/advs.202204072},
year = {2023}
}

@article{Peek2022,

 author = {Peek, Sherman E. and Ward, Jacob and Bankson, Stephen and Shah, Archit and Sellers, John A. and Adams, Mark L. and Hamilton, Michael C.},

 title = {Additive manufacturing and characterization of microstructures using two-photon polymerization for use in cryogenic applications},

 journal = {MRS Communications},

 year = {2022},

 doi = {10.1557/s43578-022-00610-5}

}

@article{Ruchka_2022,
doi = {10.1088/2058-9565/ac796c},
url = {https://doi.org/10.1088/2058-9565/ac796c},
year = {2022},
month = {jul},
publisher = {IOP Publishing},
volume = {7},
number = {4},
pages = {045011},
author = {Ruchka, Pavel and Hammer, Sina and Rockenhäuser, Marian and Albrecht, Ralf and Drozella, Johannes and Thiele, Simon and Giessen, Harald and Langen, Tim},
title = {Microscopic 3D printed optical tweezers for atomic quantum technology},
journal = {Quantum Science and Technology}
}
\bibliographystyle{spiebib} 

\end{document}